# Model checking an Epistemic $\mu$-calculus with Synchronous and Perfect Recall Semantics[*]

Rodica Bozianu
École Normale Supérieure de Cachan,
61, avenue du Pdt. Wilson,
94235 Cachan cedex

Cătălin Dima
LACL,
Université Paris Est-Créteil,
61 av. du G-ral de Gaulle,
94010 Créteil, France

Constantin Enea
LIAFA, CNRS UMR 7089,
Université Paris Diderot - Paris 7,
Case 7014, 75205 Paris Cedex 13, France

## ABSTRACT

We identify a subproblem of the model-checking problem for the epistemic $\mu$-calculus which is decidable. Formulas in the instances of this subproblem allow free variables within the scope of epistemic modalities in a restricted form that avoids embodying any form of common knowledge. Our subproblem subsumes known decidable fragments of epistemic $CTL/LTL$, may express winning strategies in two-player games with one player having imperfect information and non-observable objectives, and, with a suitable encoding, decidable instances of the model-checking problem for $ATL_{iR}$.

## 1. INTRODUCTION

The epistemic $\mu$-calculus is an enrichment of the $\mu$-calculus on trees with individual epistemic modalities $K_a$ (and its dual, denoted $P_a$). It is designed with the aim that, like the classical modal $\mu$-calculus, it would subsume most combinations of temporal and epistemic logics. The epistemic $\mu$-calculus is more expressive than linear or branching temporal epistemic logics [15, 24], propositional dynamic epistemic logics [25], or the alternating epistemic $\mu$-calculus [6]. On the other hand, some gaps in its expressive power seem to exist, as witnessed by recent observations in [6] showing that formulas like $\langle\!\langle a \rangle\!\rangle p_1 \mathcal{U} p_2$ are not expressible in the *Alternating Epistemic $\mu$-calculus*. This expressivity gap can be reproduced in the epistemic $\mu$-calculus, though the epistemic $\mu$-calculus is richer than the alternating $\mu$-calculus.

The model-checking problem for epistemic $\mu$-calculus is undecidable in the presence of a semantics with perfect recall, as it is more expressive than combinations of temporal epistemic logics that include the common knowledge operator. A rather straightforward fragment of the epistemic $\mu$-calculus which has a decidable model-checking problem is the one in which knowledge modalities apply only to closed formulas, that is, formulas in which all second-order variables are bound by some fixpoint operator. The decidability of this fragment follows from recent results on the decidability of the emptiness problem for two player games [7].

However more expressive fragments having a decidable model-checking problem seem to exist. For example, winning strategies in two-player games in which one player has imperfect information and non-observable winning conditions can be encoded as fixpoint formulas in the epistemic $\mu$-calculus, but not in the above-mentioned restricted fragment. The same holds for some formulas

[*]Work partially supported by the ANR research project "EQINOCS" no. ANR-11-BS02-0004

in $ATL$ with imperfect information and perfect recall ($ATL_{iR}$) [23, 5]: the $ATL$ formula $\langle\!\langle a \rangle\!\rangle \square p$ can be expressed in a modal $\mu$-calculus of knowledge as

$$\nu Z. \bigvee_{\alpha \in Act_a} K_a\bigl(p \wedge \bigwedge_{\beta \in Act_{Ag \setminus \{a\}}} [\alpha, \beta] Z\bigr)$$

And there are variants of $ATL_{iR}$ for which the model-checking problem is decidable [10]. Note that a translation of each instance of the model-checking problem for $ATL$ into instances of the model-checking problems for the epistemic $\mu$-calculus is also possible but requires the modification of the models.

Our aim in this paper is to identify a larger and decidable class of instances of the model-checking problem for the epistemic $\mu$-calculus. The fragment we propose here allows an epistemic modality $K_a$ to be applied to a non-closed $\mu$-calculus formula $\phi$, but in such a way that avoids expressing properties that construct any variant of common knowledge for two or more agents. Roughly, the technical restriction is the following: two epistemic operators, referring to the knowledge of two different agents $a$ and $b$, can be applied to non-closed parts of a formula only if the two agents have *compatible* observations in the system $M$ in the sense that the observability relation of one of the agents is a refinement of the observability relation of the other. Similar restrictions have been proposed for various combinations of temporal epistemic logics [12], or for the synthesis problem in distributed environments [18, 27, 13]. The variant presented here relies on a *concrete* semantics, in the sense of [9], with the observability relation for each agent $a$ being identified, in the given system $M$, by a subset $\Pi_a$ of atomic propositions. We require this in order to syntactically define our decidable subproblem: the compatibility of two observability relations $\sim_a$ and $\sim_b$ is specified by imposing that either $\Pi_a \subseteq \Pi_b$ or vice-versa.

The epistemic $\mu$-calculus with perfect recall has a history-based semantics: for each finite transition system $T$, the formulas of the epistemic $\mu$-calculus must be interpreted over the *tree unfolding* of $T$. This makes it closer with the tree interpretations of the $\mu$-calculus from [11]. For the classical $\mu$-calculus, there are two ways of proving that the satisfiability and the model-checking problem for the tree interpretation of the logic are decidable: either by providing translations to parity games, or by means of a Finite Model Theorem which ensures that a formula has a tree interpretation iff it has a *state-based* interpretation over a finite transition system (this is known to be equivalent with memoryless determinacy for parity games, see e.g. [4]).

The generalization of the automata approach does not seem to be possible for epistemic $\mu$-calculus, mainly due to the absence of an appropriate generalization of tree automata equivalent with the epistemic $\mu$-calculus. So we take the approach of providing a generalization of the Finite Model Theorem for our fragment of the epistemic $\mu$-calculus. This result says roughly that the tree interpretation of a formula over the tree unfolding of a given finite transition system $T$ which contains the epistemic operators $K_a$ or $P_a$ is exactly the "tree unfolding" of the finitary interpretation of the formula in a second transition system $T'$, which is obtained by determinizing the projection of $T$ onto the observations of agent $a$, a construction that is common for decidable fragments of temporal epistemic logics. Our contribution consists in showing that this construction can be applied for all instances in our model-checking subproblem. The proof is given in terms of commutative diagramms between boolean algebraic operators that are the interpretations of non-closed formulas.

The model checking subproblem is non-elementary hard due to the non-elementary hardness of the model-checking problem for the linear temporal logic of knowledge [26]. In the full version of this paper [3], we provide a self-contained proof of this result, by a reduction of the emptiness problem for star-free regular expressions.

The rest of the paper is divided as follows: in the next section we recall the semantics of the $\mu$-calculus and adapt it to our epistemic extension, both for the tree interpretation and the finitary interpretation. We then give, in the third section, our weak variant of the Finite Model Theorem for the classical $\mu$-calculus. The fourth section serves for introducing our fragment of the epistemic $\mu$-calculus and for proving the decidability of its model-checking problem. We end with a section with conclusions and comments.

An extended version of this paper with proofs is available as [3].

## 2. PRELIMINARIES

We start by fixing a series of notions and notations used in the rest of the paper.

$A^*$ denotes the set of words over $A$. The length of $\alpha \in A^*$, is denoted $|\alpha|$ and the prefix of $\alpha$ up to position $i$ is denoted $\alpha[1..i]$. Hence, $\alpha[1..0] = \varepsilon$ is the empty word. The prefix ordering on $A^*$ is denoted $\leq$ ($<$ for the strict prefix ordering).

Given a set $A$ and an integer $n \in \mathbb{N}$, an $A$-*tree of outdegree* $\leq n$ is a partial function $t : [1\ldots n]^* \rightharpoonup A$ whose support, denoted $\mathsf{supp}(t)$, is a prefix-closed subset of the finite sequences of integers in $[1\ldots n]$. A *node* of $t$ is an element of its support. A *path* in $t$ is a pair $(x, \rho)$ consisting of a node $x$ and the sequence of $t$-labels of all the nodes which are prefixes of $x$, $\rho = \big(t(x[1\ldots i])\big)_{0 \leq i \leq |x|}$.

**Boolean operators:** Given a set $A$, a *boolean $A$-operator* is a mapping $f : (2^A)^n \to 2^A$.

For an $A$-operator $f : (2^A)^n \to 2^A$, a tuple of sets $B_1, \ldots, B_n \subseteq A$ and some $k \leq n$ we denote $f_k(B_1, \ldots, B_{k-1}, \bullet, B_{k+1}, \ldots, B_n) : 2^A \to 2^A$ the $A$-operator with

$$f_k(B_1, \ldots, B_{k-1}, \bullet, B_{k+1}, \ldots, B_n)(B)$$
$$= f(B_1, \ldots, B_{k-1}, B, B_{k+1}, \ldots, B_n)$$

Note that when $f$ is monotone, $f_k(B_1, \ldots, B_{k-1}, \bullet, B_{k+1}, \ldots, B_n)$ is monotone too.

Following the Knaster-Tarski theorem, any monotone $A$-operator $f : 2^A \to 2^A$ has a unique least and greatest fixpoint, denoted $\mathsf{lfp}_f$, resp. $\mathsf{gfp}_f$. We may then define two $A$-operators, $\mathsf{lfp}_f^k : (2^A)^n \to 2^A$ and $\mathsf{gfp}_f^k : (2^A)^n \to 2^A$, respectively as:

$$\mathsf{lfp}_f^k(B_1, \ldots, B_n) = \mathsf{lfp}_{f_k(B_1, \ldots, B_{k-1}, \bullet, B_{k+1}, \ldots, B_n)}$$
$$\mathsf{gfp}_f^k(B_1, \ldots, B_n) = \mathsf{gfp}_{f_k(B_1, \ldots, B_{k-1}, \bullet, B_{k+1}, \ldots, B_n)}$$

Note that both these $A$-operators are constant in their $k$-th argument. It is well-known that both operators are monotone if $f$ is monotone.

## 3. THE $\mu$-CALCULUS OF KNOWLEDGE

**Syntax:** The syntax of the **epistemic $\mu$-calculus** is based on the following sets of symbols: a finite set of *agents* $Ag$, a family of sets of *atomic propositions* $(\Pi_a)_{a \in Ag}$ for which we denote $\Pi = \bigcup_{a \in Ag} \Pi_a$ and a set of *fixpoint variables* $\mathcal{Z} = \{Z_1, Z_2, \ldots\}$.

The grammar for the formulas of the epistemic $\mu$-calculus is:

$$\varphi ::= p \mid \varphi \wedge \varphi \mid \neg \varphi \mid AX\varphi \mid K_a\phi \mid \mu Z.\varphi$$

where $p \in \Pi$, $a \in Ag$ and $Z \in \mathcal{Z}$, and with the usual restriction that an operator $\mu Z$ may be applied on formulas in which the variable $Z$ has only positive occurrences.

Formulas of the type $K_a\phi$ are read as *agent $a$ knows that $\phi$ holds*. $\mu Z$ is the *least fixpoint* operator, while $AX$ is the usual *nexttime* operator from CTL, universally quantified over the successors of the current state.

Several derived operators can be defined as usual:

1. The dual of $AX$ is denoted $EX$ and defined as $EX\phi \equiv \neg AX\neg\phi$.

2. The dual of $K_a$ is denoted $P_a$ and defined as $P_a\phi \equiv \neg K_a\neg\phi$. $P_a\phi$ reads as *agent $a$ considers that $\phi$ is possible*.

3. The greatest fixpoint operator is denoted $\nu Z$ and defined as $\nu Z.\phi \equiv \neg \mu Z.\neg\phi[Z/\neg Z]$, where $\phi[Z/\neg Z]$ is the result of the syntactic substitution of each occurrence of $Z$ with $\neg Z$ in $\phi$.

As usual, for a subset of agents $A \subseteq Ag$ we may denote $E_A$ the "everybody knows" operator, $E_A\phi = \bigwedge_{a \in A} K_a\phi$.

Since our model checking construction relies heavily on formulas being interpreted as monotone mappings and, on the other side, set complementation (which is the usual interpretation of negation) is not a monotone operator we will prefer the following syntax *in positive form* for the epistemic $\mu$-calculus:

$$\varphi ::= p \mid \neg p \mid Z \mid \varphi \wedge \varphi \mid \varphi \vee \varphi \mid AX\varphi \mid EX\varphi \mid$$
$$K_a\phi \mid P_a\phi \mid \mu Z.\varphi \mid \nu Z.\varphi$$

It is easy to see that each formula of the epistemic $\mu$-calculus can be transformed into a formula in positive form, by pushing negations through the operators and using the definitions of the dual operators.

The fragment of the epistemic $\mu$-calculus which does not involve the knowledge operator $K_a$ (or its dual $P_a$) is called here the *plain $\mu$-calculus*, or simply the $\mu$-calculus, when there's no risk of confusion. As usual, we say that a formula $\phi$ is *closed* if each variable $Z$ in $\phi$ occurs in the scope of a fixpoint operator for $Z$.

We will also briefly consider in this paper the *modal epistemic $\mu$-calculus*, for the sake of comparison with other combinations of temporal and epistemic logics. The language of this variant of the epistemic $\mu$-calculus is based on a family of sets $(Act_a)_{a \in Ag}$, meant to represent actions available to each agent at a given state. Its grammar is the following:

$$\varphi ::= p \mid \varphi \wedge \varphi \mid \neg\varphi \mid \langle \alpha \rangle \varphi \mid K_a\phi \mid \mu Z.\varphi$$

where $p \in \Pi$, $a \in Ag$, $\alpha \in \times_{a \in Ag} Act_a$ and $Z \in \mathcal{Z}$, and bearing the same restriction on the utilization of the fixpoint operators. Formulas of the type $\langle \alpha \rangle \varphi$ read as *there exists an $\alpha$-successor of the current state in which $\varphi$ holds*. The dual of the $\langle \alpha \rangle$ operator is denoted $[\alpha]$.

### 3.1 Semantics

**The tree semantics** of the epistemic $\mu$-calculus is given in terms of $2^{\Pi \cup \mathcal{Z}}$-trees endowed with a family of relations $(\sim_a)_{a \in Ag}$ with $\sim_a \subseteq \mathsf{supp}(t) \times \mathsf{supp}(t)$. The nodes of the tree represent instant descriptions of the system state, while the relation $\sim_a$ models the *indistinguishability* relation which disallows agent $a$ to tell apart two behaviors of the system.

Formaly, given a tree $t$ and the family of relations $(\sim_a)_{a \in Ag}$, each formula $\phi$ which contains variables $Z_1, \ldots, Z_n$ is associated with a $\mathsf{supp}(t)$-operator $\|\phi\| : \left(2^{\mathsf{supp}(t)}\right)^n \to 2^{\mathsf{supp}(t)}$ defined by structural induction, as follows:

- The atom $p$ is interpreted as the constant $\mathsf{supp}(t)$-operator $\|p\| : \left(2^{\mathsf{supp}(t)}\right)^n \to 2^{\mathsf{supp}(t)}$ defined as follows:

  $$\|p\|(S_1, \ldots, S_n) = \{x \in \mathsf{supp}(t) \mid p \in \pi(t(x))\}$$

- The semantics of the boolean operators is classical:

  $$\|\neg\phi\| = \mathsf{supp}(t) \smallsetminus \|\phi\|$$
  $$\|\phi_1 \wedge \phi_2\| = \|\phi_1\| \cap \|\phi_2\|$$

- Each variable $Z_i \in \mathcal{Z}$ is interpreted as the $i$-th projection on $\left(2^{\mathsf{supp}(t)}\right)^n$, that is, as the operator $\|Z_i\| : \left(2^{\mathsf{supp}(t)}\right)^n \to 2^{\mathsf{supp}(t)}$ with

  $$\|Z_i\|(S_1, \ldots, S_n) = S_i, \forall S_1, \ldots, S_n \subseteq \mathsf{supp}(t)$$

- The nexttime operator $AX$ is mapped to a $\mathsf{supp}(t)$-operator, denoted $AX : 2^{\mathsf{supp}(t)} \to 2^{\mathsf{supp}(t)}$, such that for each $S \subseteq \mathsf{supp}(t)$,

  $$AX(S) = \{x \in \mathsf{supp}(t) \mid \forall i \in \mathbb{N} \text{ if } xi \in \mathsf{supp}(t) \text{ then } xi \in S\}$$

  Then the semantics of formulas of the type $AX\phi$ is defined as:

  $$\|AX\phi\| = AX \circ \|\phi\|$$

- Each epistemic operator $K_a$ is mapped to a $\mathsf{supp}(t)$-operator denoted $K_a : 2^{\mathsf{supp}(t)} \to 2^{\mathsf{supp}(t)}$, such that for each $S \subseteq \mathsf{supp}(t)$,

  $$K_a(S) = \{x \in \mathsf{supp}(t) \mid \forall y \in \mathsf{supp}(t), \text{ if } x \sim_a y \text{ then } y \in S\}$$

  Then the semantics of formulas of the type $K_a\phi$ is defined as:

  $$\|K_a\phi\| = K_a \circ \|\phi\|$$

- The fixpoint operators are interpreted as usual:

  $$\|\mu Z_i.\phi\| = \mathsf{lfp}^i_{\|\phi\|}$$

We denote $t \vDash \phi$ iff $\varepsilon \in \|\phi\|$.

The semantics of the epistemic $\mu$-calculus can be also described without set complementation, by keeping the definition of negation only for atomic formulas, and appending the following definitions:

$$\|\neg p\|(S_1, \ldots, S_n) = \{x \in \mathsf{supp}(t) \mid p \notin \pi(t(x))\}$$
$$EX(S) = \{x \in \mathsf{supp}(t) \mid \exists i \in \mathbb{N} \text{ with } xi \in \mathsf{supp}(t) \text{ and } xi \in S\}$$
$$\|EX\phi\| = EX \circ \|\phi\|$$
$$P_a(S) = \{x \in \mathsf{supp}(t) \mid \exists y \in \mathsf{supp}(t) \text{ with } x \sim_a y \text{ and } y \in S\}$$
$$\|P_a\phi\| = P_a \circ \|\phi\|$$
$$\|\nu Z_i.\phi\| = \mathsf{gfp}^i_{\|\phi\|}$$

Note that, this way, all operators are interpreted as monotone $\mathsf{supp}(t)$-operators, which is more convenient for manipulating fixpoints.

As we are interested in the model-checking problem, we will only work with finitely-generated trees as models for the epistemic $\mu$-calculus. These finitely-generated models occur as unfoldings of *multi-agent systems*, whose definition is recalled here.

A **multi-agent system** is a tuple $M = \left(Q, Ag, \delta, q_0, \Pi, (\Pi_a)_{a \in Ag}, \pi\right)$ with $Ag$ being the set of agents, $Q$ the set of states, $q_0$ the initial state of the system, $\delta \subseteq Q \times Q$, $\Pi$ the set of *atomic propositions*, $\pi : Q \to 2^\Pi$ is the *state labeling* and for all $a \in Ag$, $\Pi_a \subseteq \Pi$ is the set of atoms *observable by* agent $a$. A run in the system $M$ is an infinite sequence of states $\rho = q_0 q_1 q_2 \ldots$ such that $(q_i, q_{i+1}) \in \delta$ for all $i \geq 0$. The set of finite runs in $M$ is denoted $\mathsf{Runs}(M)$. Throughout this paper we consider only finite systems, with $Q = \{1, \ldots, n\}$ and $q_0 = 1$, and we assume that $Q$ contains only reachable states.

The $Q$-tree representing the *unfolding* of a multi-agent system $M$ is denoted $t_M$ and defined by

$$\mathsf{supp}(t_M) = \{x \in \mathbb{N}^* \mid 1x \in \mathsf{Runs}(M)\} \text{ and } t_M(x) = x[|x|]$$

The actual tree that can be used as a model of the epistemic $\mu$-calculus is $\pi(t_M) = \pi \circ t_M : \mathsf{supp}(t_M) \to 2^\Pi$. We denote this tree as $\pi t_M$.

The family of indistinguishability relations $(\sim_a)_{a \in Ag}$ that we consider in this paper are defined as follows: for any two positions $x, y \in \mathsf{supp}(t_M)$ with $|x| = |y|$, we denote $x \sim_a y$ if for any $n \leq |x|$ we have that

$$\pi(t(x[1..n])) \cap \Pi_a = \pi(t(y[1..n])) \cap \Pi_a$$

This way, the indistinguishability relation $\sim_a$ models the fact that agent $a$ has perfect knowledge of the absolute time and remembers all his past observations – that is, $\sim_a$ is a *synchronous and perfect recall* indistinguishability.

*Definition 1.* The **model-checking problem** for the epistemic $\mu$-calculus is the problem of deciding, given a multi-agent system $M$ and a closed formula $\phi$, whether $\pi t_M \vDash \phi$.

The undecidability of the model-checking problem for combinations of temporal and epistemic logics based on a synchronous and perfect recall semantics and containing the common knowledge operator [26, 25], together with the connections between the epistemic $\mu$-calculus and such temporal epistemic logics that are explored in the next section, imply the following result:

THEOREM 1. *The model-checking problem for the epistemic $\mu$-calculus is undecidable.*

The semantics of the **modal epistemic $\mu$-calculus** is a slight variation of the above semantics, in that we utilize a different type of trees, as mappings $t : \mathbb{N} \to 2^{\Pi \cup \mathcal{Z}} \times \times_{a \in Ag} Act_a$. We decompose such a tree as $t = (t^{node}, t^{edge})$: the tree of *nodes* is $t^{node}(x) = t(x)\big|_{2^{\Pi \cup \mathcal{Z}}}$, while the tree of *edges* is $t^{edge}(x)t(x)\big|_{\times_{a \in Ag} Act_a}$. The only item that changes in the above list of semantic rules for operators is that we replace the definition of the nextttime operator with the following definition of the a boolean operator $\langle \alpha \rangle : 2^{\mathsf{supp}(t)} \to 2^{\mathsf{supp}(t)}$: for each $S \subseteq \mathsf{supp}(t)$,

$$\langle \alpha \rangle(S) = \{x \in \mathsf{supp}(t) \mid \exists i \in \mathbb{N} \text{ with } xi \in \mathsf{supp}(t) \text{ and } xi \in S\}$$

A family of indistinguishability relations in such a tree model for the modal epistemic $\mu$-calculus is, like in the non-modal case, a family of relations $(\sim_a)_{a \in Ag}$ with $\sim_a \subseteq \mathsf{supp}(t) \times \mathsf{supp}(t)$.

Then, finite presentations of tree models for the modal epistemic $\mu$-calculus are *multi-agent systems with transition labels*, which are tuples $M = \left(Q, Ag, (Act_a)_{a \in Ag}, \delta, q_0, \Pi, (\Pi_a)_{a \in Ag}, \pi\right)$ with $\delta \subseteq Q \times \times_{a \in Ag} Act_a \times Q$ and all the other components bearing the same name and definition as in (plain) multi-agent systems.

The tree representing the *unfolding* of $M$, denoted $t_M$ also, is defined inductively as follows:

- $\varepsilon \in \mathsf{supp}(t_M)$ and $t^{node}(\varepsilon) = q_0$; $t^{edge}(\varepsilon)$ is left unconstrained.

- If $x \in \mathsf{supp}(t_M)$ and $t^{node}(x) = q$, then for each state $r$ and tuple of actions $\alpha \in \times_{a \in Ag} Act_a$, for which $q \xrightarrow{\alpha} r \in \delta$, there exists a succesor of $x$ denoted $xi_{r,\alpha}$, and $t(xi_{r,\alpha}) = (r, \alpha)$.

- All successors of a node $x$ are obtained by the previous rule.

The family of indistinguishability relations is defined in a slightly different way for unfoldings of transition-labeled multi-agent systems, as agents may know their own past actions. Formally, for two nodes $x, y \in \mathsf{supp}(t_M)$ and an agent $a \in Ag$ we put $x \sim_a y$ if for any $n \leq |x|$ we have that

$$\pi(t^{node}(x[1..n])) \cap \Pi_a = \pi(t^{node}(y[1..n])) \cap \Pi_a$$
$$t^{edge}(x[1..n])\big|_a = t^{edge}(y[1..n])\big|_a$$

The modal epistemic $\mu$-calculus can be translated into the (non-modal) epistemic $\mu$-calculus by converting each action name $\alpha_a \in Act_a$ into an atomic proposition, so the main results of this paper generalize easily to this calculus.

### 3.2 Comparison with other temporal epistemic frameworks

In this subsection we discuss the relationship between the epistemic $\mu$-calculus and other temporal epistemic logics or game models with imperfect information and perfect recall.

First, it is easy to see that the epistemic $\mu$-calculus is more expressive than linear or branching temporal epistemic logics with common knowledge operators [15]. This was already noted e.g. in [24], since the following fixpoint formula defines the common knowledge operator for two agents: $C_{a,b}\phi = \nu Z.(\phi \wedge K_a Z \wedge K_b Z)$.

Secondly, the (modal variant of the) epistemic $\mu$-calculus is more expressive than the alternating epistemic $\mu$-calculus of [6], due to the possibility to insert knowledge operators "in between" the quantifiers that occur in the semantics of the coalition operators. More precisely, for any instance of the model-checking problem for the alternating epistemic $\mu$-calculus, let $Act_A$, denote, for each set of agents $A \subseteq Ag$, the cartesian product of the set of action symbols for each agent in $A$, $Act_A = \times_{a \in A} Act_a$. Then:

$$\langle\!\langle A \rangle\!\rangle X\phi = \bigvee_{\alpha \in Act_A} \big( K_a \bigwedge_{\beta \in Act_{Ag \setminus A}} [\alpha, \beta]\phi \big)$$
$$[\![ A ]\!] X\phi = \bigwedge_{\alpha \in Act_A} \big( P_a \bigvee_{\beta \in Act_{Ag \setminus A}} [\alpha, \beta]\phi \big)$$

Recall briefly that the *strategy operator* $\langle\!\langle A \rangle\!\rangle \phi$ says that the agents in the group (coalition) $A$ have a *strategy* ensuring that, whatever the other agents do, the objective $\phi$ is achieved on each resulting run. Also the strategy must be based on the observability of each agent of the system state. See [5] for a recent account on alternating temporal logics.

The relationship with $ATL_{iR}$ is more involved, as we detail in the sequel. Formulas of the type $\langle\!\langle A \rangle\!\rangle \square p$ can be expressed as the fixpoint formula $\nu Z. \bigvee_{\alpha \in Act_a} K_a \big( p \wedge \bigwedge_{\beta \in Act_{Ag \setminus \{a\}}} [\alpha, \beta]Z \big)$.

On the other hand, formulas containing the until operator cannot be translated into the epistemic $\mu$-calculus. The reason is explained in [6]: in formulas of the type $\langle\!\langle a \rangle\!\rangle \Diamond p$ the objective $p$ might not be observable by the agent $a$, who might only be able to know, in the future of some given time instant, that sometimes in the past of that future moment (but after the reference instant), the objective was achieved on all identically observable traces.

However, for the decidable case of coalitions based on distributed knowledge [10], a translation exists for each instance of the model-checking problem. We provide here this translation for simple reachability formulas: given an $ATL_{iR}$ formula $\phi = \langle\!\langle a \rangle\!\rangle p_1 \mathcal{U} p_2$, a multi-agent system $M$ and a finite run $\rho$ in $M$, the instance of the model-checking problem $M, \rho \vDash \phi$ can be translated to an instance of the model-checking problem in the epistemic modal $\mu$-calculus of the following formula:

$$\tilde{\phi} = \mu Z. \bigvee_{\alpha \in Act_a} K_a \big( p_2 \vee past_{p_2} \vee \big( p_1 \wedge \bigwedge_{\beta \in Act_{Ag \setminus \{a\}}} [\alpha, \beta]Z \big) \big) \quad (1)$$

and the *modified* system $M'$, in which the new atomic proposition $past_{p_2}$ labels all the states occurring *after* a state carrying a $p_2$ and lying on runs which extend $\rho$. This mechanism is similar with the "bookkeeping" employed in the two-player games utilized in [10] for checking whether the same formula $\phi$ holds at a state of a multi-agent system.

Formally, given a multi-agent system
$M = (Q, Ag, \delta, q_0, \Pi, (\Pi_a)_{a \in Ag}, \pi, (Act_a)_{a \in Ag})$, we build the system $M' = (Q', Ag, \delta', q'_0, \Pi, (\Pi_a)_{a \in Ag}, \pi', (Act'_a)_{a \in Ag})$ in which:

- $Q' = Q \times \{0, 1\}$ and $q'_0 = (q_0, 0)$.

- $\pi'(q, 0) = \pi(q)$, $\pi'(q, 1) = \pi(q) \cup \{past_{p_2}\}$.

- $Act'_{a_0} = Act_{a_0} \times \{0, 1\}$ and $Act'_b = Act_b$ for all $b \neq a_0$.

- For any transition $q \xrightarrow{(\alpha, \beta)} r$ with $\alpha \in Act_{a_0}$ and $\beta = (\beta_b)_{b \neq a_0}$, we put in $\delta'$ the following transitions:

  - $(q, 0) \xrightarrow{((\alpha, 0), \beta)} (r, 0)$
  - $(q, 1) \xrightarrow{((\alpha, x), \beta)} (r, 1)$, $x \in \{0, 1\}$
  - $(q, 0) \xrightarrow{((\alpha, 1), \beta)} (r, 1)$ if $p_2 \in \pi(q)$
  - $(q, 0) \xrightarrow{((\alpha, 1), \beta)} (r, 0)$ if $p_2 \notin \pi(q)$

Note that, given a node $x \in \mathsf{supp}(t_{M'})$, if we replace, on the path from the root to $x$, all actions of the type $(\alpha, 0)$ with $\alpha$, we get a run in $t_M$ corresponding with a note of $t_M$. We denote this corresponding node as $x\big|_M$. Furthermore, for each node $x \in \mathsf{supp}(t_M)$, we denote $x \uparrow^{M'}$ the node in $\mathsf{supp}(t_{M'})$ with $(x \uparrow^{M'})\big|_M = x$ and having the property that on the path from the root of $t_{M'}$ to $x \uparrow^{M'}$, $a$'s actions are only of the type $(\alpha, 1)$.

The following proposition gives the connection between the instances of the model-checking problem in $M$ and $M'$:

PROPOSITION 2. *For each node $x$ in the tree $t_M$, $x \vDash \phi = \langle\!\langle a_0 \rangle\!\rangle p_1 \mathcal{U} p_2$ if and only if $x \uparrow^{M'} \vDash \tilde{\phi}$, with $\tilde{\phi}$ defined as follows:*

$$\tilde{\phi} = \mu Z. \bigvee_{\alpha \in Act_{a_0}} K_{a_0} \big( p_2 \vee past_{p_2} \vee \big( p_1 \wedge \bigwedge_{\beta \in Act_{Ag \setminus \{a_0\}}} [\alpha, \beta]Z \big) \big)$$

*The same property holds for $\phi = [\![ a_0 ]\!] p_1 \mathcal{U} p_2$ (which reads "agent $a_0$ cannot avoid $p_1 \mathcal{U} p_2$") and*

$$\tilde{\phi} = \mu Z. \bigwedge_{\alpha \in Act_{a_0}} P_{a_0} \big( p_2 \vee past_{p_2} \vee \big( p_1 \wedge \bigvee_{\beta \in Act_{Ag \setminus \{a_0\}}} \langle \alpha, \beta \rangle Z \big) \big)$$

The problem of *solving multi-player games with imperfect information* can also be translated into the epistemic $\mu$-calculus. Recall that a (synchronous) two-player game is a tuple

$$G = \big( Q, Act_0, Act_1, \delta, Q_0, Obs_0, Obs_1, o_0, o_1, par \big)$$

with $Q$ denoting the set of states, $Act_0$ (resp $Act_1$) denoting the set of actions available to player 0 (resp. player 1), $\delta \subseteq Q \times Act_0 \times Act_1 \times Q$ denoting the transition relation, $Obs_0$, resp. $Obs_1$ denoting finite sets of observations available to agent 0 (resp. agent 1), $o_0 : Q \to Obs_0$, resp. $o_1 : Q \to Obs_1$ denoting the observability relation for each player and $par : Q \to \mathbb{N}$ defining the *priority* of each state.

A player $i$ ($i \in \{0, 1\}$) plays by choosing a *feasible strategy*, which is a mapping $\sigma : (Obs_i)^* \to Act_i$. A strategy for $i$ is *winning* when each runs that is compatible with that strategy satisfies the property that the maximal priority of a state which occurs infinitely often in the run is even. The winning condition might be non-observable to player $i$, as there might exist states $q_1, q_2 \in Q$ that are identically observable by player $i$, i.e. $o_i(q_1) = o_i(q_2)$, might have different priorities.

The set of winning strategies for a player in a multi-player game with imperfect information is then expressible within the epistemic $\mu$-calculus, similarly to the encoding of the set of winning strategies in a parity game into the $\mu$-calculus from e.g. [11, 22]. Assuming that the largest priority in $Q$ is even and the atomic proposition $p_k$ holds exactly in all states with priority $k$, the following epistemic modal $\mu$-calculus formula encodes the winning strategies for player $i$:

$$\nu Z_n \mu Z_{n-1} \ldots \mu Z_1. \bigvee_{\alpha \in Act_i} K_a \bigvee_{k \leq n} \left( p_k \wedge \bigwedge_{\beta \in Act_{1-i}} [\alpha, \beta] Z_k \right)$$

provided that player $i$'s indistinguishability in the multi-agent system constructed from $G$ is based on $Obs_i$.

### 3.3 Revisiting the decidability of the model checking problem for the tree semantics of the plain $\mu$-calculus

In this subsection we provide a variant of the Finite Model Theorem for the $\mu$-calculus, which will serve as a basis for our search of a decidable subproblem of the model-checking problem for the epistemic $\mu$-calculus.

Given a multi-agent system $M = (Q, Ag, \delta, q_0, \Pi, (\Pi_a)_{a \in Ag}, \pi)$, and an agent $a \in Ag$, we define the relation $\Gamma_a^M \subseteq Q \times Q$ as follows: $(q, r) \in \Gamma_a^M$ if for any run $\rho$ in $M$ ending in $q$ (i.e. $\rho[|\rho|] = q$) there exists a run $\rho'$ ending in $r$ with $\rho \sim_a \rho'$.

We now define a second semantics for the epistemic $\mu$-calculus, which works on the *set of states* of a multi-agent system $M$, necessary for the decision problem. This semantics is the extension of the state-based semantics for the $\mu$-calculus [21] by defining a state-based semantics for the epistemic operators.

Formally, each formula $\phi$ which contains variables $Z_1, \ldots, Z_n$ is associated with a $Q$-operator $\lceil \phi \rceil_M : (2^Q)^n \to 2^Q$, again by structural induction (we provide here the semantics for the epistemic $\mu$-calculus in positive form):

- $\lceil p \rceil_M$ resp. $\lceil \neg p \rceil_M$ are the constant $Q$-operators
$$\lceil p \rceil_M(S_1, \ldots, S_n) = \{q \in Q \mid p \in \pi(q)\}$$
$$\lceil \neg p \rceil_M(S_1, \ldots, S_n) = \{q \in Q \mid p \notin \pi(q)\}$$

- $\lceil Z_i \rceil_M : (2^Q)^n \to 2^Q$ is the $i$-th projection $Q$-operator, i.e. given $S_1, \ldots, S_n \subseteq Q$, $\lceil Z_i \rceil_M(S_1, \ldots, S_n) = S_i$.

- $\lceil \phi_1 \vee \phi_2 \rceil_M = \lceil \phi_1 \rceil_M \cup \lceil \phi_2 \rceil_M$, and $\lceil \phi_1 \wedge \phi_2 \rceil_M = \lceil \phi_1 \rceil_M \cap \lceil \phi_2 \rceil_M$.

- Both nexttime modalities are associated with $Q$-operators $AX^f, EX^f : 2^Q \to 2^Q$ such that:
$$\lceil AX\phi \rceil_M = AX^f \circ \lceil \phi \rceil, \qquad \lceil EX\phi \rceil_M = EX^f \circ \lceil \phi \rceil$$

where:
$$AX^f(S) = \{q \in Q \mid \forall r \in Q \text{ if } (q, r) \in \delta \text{ then } r \in S\}$$
$$EX^f(S) = \{q \in Q \mid \exists r \in Q \text{ with } (q, r) \in \delta \text{ and } r \in S\}$$

- Each pair of epistemic operators $K_a/P_a$ is associated with a pair of $Q$-operators $K_a^f, P_a^f : 2^Q \to 2^Q$ such that:
$$\lceil P_a \phi \rceil_M = P_a^f \circ \lceil \phi \rceil$$
$$\lceil K_a \phi \rceil_M = K_a^f \circ \lceil \phi \rceil$$

where:
$$K_a^f(S) = \overline{\Gamma_a(\overline{S})} = \{q \in Q \mid \forall s \in Q, \text{ if } (s, q) \in \Gamma_a \text{ then } s \in S\}$$
$$P_a^f(S) = \Gamma_a(S) = \{q \in Q \mid \exists s \in S \text{ s.t. } (s, q) \in \Gamma_a\}$$

- $\lceil \mu Z_i. \phi \rceil_M = \mathsf{lfp}^i_{\lceil \phi \rceil_M}$ and $\lceil \nu Z_i. \phi \rceil_M = \mathsf{gfp}^i_{\lceil \phi \rceil_M}$.

In the sequel, when the multi-agent system $M$ is fixed, we will utilize the notation $\lceil \varphi \rceil$ instead of $\lceil \varphi \rceil_M$.

The following result, giving the connection between the tree semantics and the state-based semantics for the $\mu$-calculus, contains the essence of the Finite Model Theorem for $\mu$-calculus. The result is proved by structural induction on the formula $\phi$ in [3]:

THEOREM 3. *Given a multi-agent system $M = (Q, Ag, \delta, q_0, \Pi, (\Pi_a)_{a \in Ag}, \pi)$ in which $Q = \{1, \ldots, n\}$ and $q_0 = 1$, and a (plain) $\mu$-calculus formula $\phi$, the following diagram[1] commutes:*

$$\begin{array}{ccc} (2^Q)^n & \xrightarrow{\lceil \phi \rceil} & 2^Q \\ {\scriptstyle (t_M^{-1})^n} \downarrow & & \downarrow {\scriptstyle t_M^{-1}} \\ (2^{\mathsf{supp}(t_M)})^n & \xrightarrow{\|\phi\|} & 2^{\mathsf{supp}(t_M)} \end{array} \quad (2)$$

We also say that the diagram 2 holds (or commutes) for the formula $\phi$ in the system $M$.

The commutativity of diagram 2 is based on some commutativity properties for the tree operators and the state operators associated with all the logical operators of the $\mu$-calculus. For instance, the $AX$ operator satisfies the following commutativity property:

$$\begin{array}{ccc} 2^Q & \xrightarrow{AX^f} & 2^Q \\ {\scriptstyle (t_M^{-1})^n} \downarrow & & \downarrow {\scriptstyle t_M^{-1}} \\ 2^{\mathsf{supp}(t_M)} & \xrightarrow{AX} & 2^{\mathsf{supp}(t_M)} \end{array} \quad (3)$$

Our search will be directed towards finding particular instances of the model-checking problem where similar commutative diagrams can be provided for the epistemic operators involved in the given epistemic $\mu$-calculus formula.

## 4. A FRAGMENT OF THE EPISTEMIC $\mu$-CALCULUS WITH A DECIDABLE MODEL CHECKING PROBLEM

In this section, we first introduce some additional notations and notions. Given a multi-agent system $M$ and two agents $a_1, a_2 \in Ag$, we say that the two agents **have compatible observability** if either $\Pi_{a_1} \subseteq \Pi_{a_2}$ or $\Pi_{a_1} \supseteq \Pi_{a_2}$.

---
[1]The category in which this diagram holds is $Set$, the category of sets.

180

Given a formula $\phi$, let $T_\phi$ denote the syntactic tree of $\phi$. The following fixes the definition of $T_\phi$ by structural induction, as it will be needed in the rest of the proof. Note that, in our definition of $T_\phi$, each node labeled with a *variable* also has a *successor*, labeled with ⊤. This convention brings the property that each node in $T_\phi$ whose formula is a variable has a closed subformula (which is ⊤):

- $\text{supp}(T_p) = \{\epsilon\}$, $T_p(\epsilon) = p$,
- $\text{supp}(T_{\neg p}) = \{\epsilon\}$, $T_{\neg p}(\epsilon) = \neg p$,
- $\text{supp}(T_Z) = \{\epsilon, 1\}$, $T_Z(\epsilon) = Z$, $T_Z(1) = \top$,
- $\text{supp}(T_{Op\phi_1}) = \{\epsilon\} \cup \{1x \mid x \in \text{supp}(T_{\phi_1})\}$, $T_{Op\phi_1}(\epsilon) = Op$, $T_{Op\phi_1}(1x) = T_{\phi_1}(x)$, where $Op \in \{AX, EX, K_a, P_a, \mu Z, \nu Z\}$
- $\text{supp}(T_{\phi_1 Op \phi_2}) = \{\epsilon\} \cup \{1x \mid x \in \text{supp}(T_{\phi_1})\} \cup \{2x \mid x \in \text{supp}(T_{\phi_2})\}$, $T_{\phi_1 Op \phi_2}(\epsilon) = Op$, $T_{\phi_1 Op \phi_2}(1x) = T_{\phi_1}(x)$, $T_{\phi_1 Op \phi_2}(2x) = T_{\phi_2}(x)$, $Op \in \{\wedge, \vee\}$.

We then denote $form(x)$ the subformula of $\phi$ whose syntactic tree is $T_\phi|_x$, i.e. the subtree of $T_\phi$ rooted at $x$, and say that $x$ is **closed** if $form(x)$ is closed.

We then say that an epistemic operator $Op \in \{K_a, P_a \mid a \in Ag\}$ is **non-closed** at a node $x$ in a formula $\phi$ if $form(x)$ is not closed, $Op$ labels a node $y \geq x$ and for all the nodes $y'$ lying on the path between $x$ and $y$ we have that $form(y')$ is not closed.

For each node $x \in \text{supp}(T_\phi)$, we also define $AgNCl_\phi(x)$ as the set of agents $a$ for which $K_a$ or $P_a$ is not closed at $x$. In addition, given two distinct nodes $x_1 \prec x_2$ with $x_2$ being closed, we say that $x_2$ is a *nearest closed successor* of $x_1$ if no other closed node lies on the path from $x_1$ to $x_2$.

*Definition 2.* A formula $\phi$ is said to **mix observations of agents** $a$ **and** $b$ (or also: agents $a, b$ have **mixed observations** in $\phi$) if the following property holds

For some epistemic operators $Op_a \in \{K_a, P_a\}$, $Op_b \in \{K_b, P_b\}$ there exists a node $x$ of $T_\phi$ such that both $Op_a$ and $Op_b$ are not closed at $x$.

The **non-mixing model-checking problem** for the epistemic $\mu$-calculus is the problem of deciding whether $t_M \vDash \phi$ for a given multi-agent system $M$ and a closed formula $\phi$ bearing the restriction that any two agents $a, b$ which have mixed observations in $\phi$ have compatible observability in $M$.

All instances of the model-checking problem for $KB_n$ [15, 16], that is, $CTL$ with individual knowledge operators, are formulas of the $\mu$-calculus of non-mixing epistemic fixpoints. Other instances of this model-checking problem consist of the following formulas

$$\mu Z_1.(p \vee K_a(EX.Z_1) \wedge \nu Z_2.(q \wedge Z_1 \wedge K_a(EXZ_2)))$$
$$\mu Z_1.(p \vee K_a(EX.Z_1) \wedge \nu Z_2.(q \wedge Z_1 \wedge K_b(EXZ_2)))$$

in pair with systems $M$ in which $\Pi_a \subseteq \Pi_b$. Also any instance of the model-checking problem for the following common knowledge formula:

$$C_{a,b}\phi = \nu Z.(\phi \wedge K_a Z \vee K_b Z)$$

and with systems $M$ in which $a$ and $b$ do not have compatible observability, is not an instance of the non-mixing model-checking problem.

THEOREM 4. *The non-mixing model-checking problem for the epistemic $\mu$-calculus is decidable.*

The crux of the proof relies on a commutativity property relating $t_M^{-1}$ with the operators $K_a/K_a^f$, resp. $P_a/P_a^f$, similar with the properties relating $t_M^{-1}$ with $AX/AX^f$ in diagram 3. Unfortunately, such a commutativity property does not hold for $K_a$ in any multi-agent system $M$, as is shown in the following example.

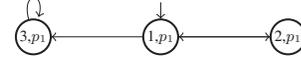

**Figure 1: A one-agent system with $\Pi_a = \{p_1\}$.**

EXAMPLE 5. *For the one-agent system in Fig. 1 we have that* $t_M^{-1}(K_a^f(\{1,3\})) \neq K_a(t_M^{-1}(\{1,3\}))$, *as*

$$t_M^{-1}(K_a^f(\{1,3\})) = \{x \in \text{supp}(t_M) \mid x[|x|] = 1\}$$
$$K_a(t_M^{-1}(\{1,3\})) = \{x \in \text{supp}(t_M) \mid x[|x|] = 1 \vee$$
$$(x[|x|] = 3 \wedge |x| \text{ is odd})\}.$$

*Definition 3.* Given two multi-agent systems $M_i = (Q_i, Ag, \delta_i, q_0^i, \Pi, (\Pi_a)_{a \in Ag}, \pi_i)$ ($i = 1, 2$) over the same set of atomic propositions, we say that $M_1$ is an **in-splitting** of $M_2$ if there exists a surjective mapping with $\chi : Q_1 \to Q_2$, satisfying the following properties:

1. For each $q, r \in Q_1$, if $(q, r) \in \delta_1$ then $(\chi(q), \chi(r)) \in \delta_2$. Moreover, for any $(q', r') \in \delta_2$ there exist $(q, r) \in \delta_1$ such that $\chi(q) = q', \chi(r) = r'$.

2. For each $q \in Q_1$, $\pi_2(\chi(q)) = \pi_1(q)$.

3. For each $q \in Q_1$, $\text{outdeg}(\chi(q)) = \text{outdeg}(q)$, where $\text{outdeg}(q)$ is the number of transitions leaving $q$.

4. $\chi(q_0^1) = q_0^2$.

The in-splitting is an **isomorphism** whenever $\chi$ is a bijection.

We will call the mapping $\chi$ as an *in-splitting mapping*. Also, we write $\chi : M_1 \xrightarrow{Ins} M_2$ to denote the fact that $\chi$ is a witness for $M_1$ being an in-splitting of $M_2$.

Note that an in-splitting mapping (term borrowed from symbolic dynamics [19]) represents a surjective functional bisimulation between two transition systems. The following proposition can be seen as a generalization of this remark (the proof is given in [3]):

PROPOSITION 6. *Consider two multi-agent systems $M_i = (Q_i, Ag, \delta_i, q_0^i, \Pi, (\Pi_a)_{a \in Ag}, \pi_i)$ ($i = 1, 2$) over the same set of atoms, connected by an in-splitting mapping $\chi : M_1 \xrightarrow{Ins} M_2$. Then for any plain $\mu$-calculus formula $\phi$ the following diagram commutes:*

$$\begin{array}{ccc} (2^{Q_1})^n & \xrightarrow{[\phi]_{M_1}} & 2^{Q_1} \\ (\chi^{-1})^n \uparrow & & \uparrow \chi^{-1} \\ (2^{Q_2})^n & \xrightarrow{[\phi]_{M_2}} & 2^{Q_2} \end{array} \quad (4)$$

REMARK 7. *Proposition 6 does not hold for any epistemic $\mu$-calculus formula. To see this, consider the system depicted in Fig. 2, which is an in-splitting of the system from Fig. 1, obtained by splitting state 3 in Fig. 1 in two states, denoted 3 and 4 in Fig. 2, (i.e. $\chi(1) = 1, \chi(2) = 2, \chi(3) = \chi(4) = 3$) with transitions $(3, 4) \in \delta$ and $(4, 4) \in \delta$.*

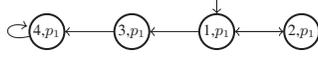

**Figure 2: An in-splitting of the system from Fig. 1.**

Note that we have
$$\lceil K_a^f X \rceil_{M_2}(\{1,2,3\}) = \{1,3\}$$
$$\lceil K_a^f X \rceil_{M_1}(\{1,2,3\}) = \{1,2,3\}$$

and hence $\lceil K_a^f X \rceil_{M_1} \circ \chi^{-1} \neq \chi^{-1} \circ \lceil K_a^f X \rceil_{M_2}$.

The following notion corresponds with the "subset construction" used for model-checking LTLK/CTLK [26, 8] or solving 2-player parity games with one player having incomplete information [7]:

*Definition 4.* Given a multi-agent system $M = (Q, Ag, \delta, q_0, \Pi, (\Pi_a)_{a \in Ag}, \pi)$, we define the multi-agent system
$$\Delta_a^{pre}(M) = (\tilde{Q}^{pre}, Ag, \tilde{\delta}, \tilde{q}_0, \Pi, (\Pi_a)_{a \in Ag}, \tilde{\pi})$$
as follows:

- $\tilde{Q}^{pre} = \{(s,S) \mid s \in Q, S \subseteq \{q \in Q \mid \pi_a(q) = \pi_a(s)\}\}$ and $\tilde{q}_0 = (q_0, \{q_0\})$.
- $\tilde{\delta}$ is composed of all tuples of the form $((s,S),(r,R))$ where $(s,r) \in \delta$ and $R = \{r' \in Q \mid \pi_a(r') = \pi_a(r) \text{ and } \exists s' \in S \text{ with } (s',r') \in \delta\}$.
- $\tilde{\pi}(s,S) = \pi(S) = \pi(s)$.

The $a$-**distinction** of $M$, denoted $\Delta_a(M)$, is the restriction of $\Delta_a^{pre}(M)$ to reachable states, i.e.,
$$\Delta_a(M) = (\tilde{Q}, Ag, \tilde{\delta}|_{\tilde{Q}}, \tilde{q}_0, \Pi, (\Pi_a)_{a \in Ag}, \tilde{\pi}|_{\tilde{Q}})$$
where $\tilde{Q} = \{\tilde{s} \in \tilde{Q}^{pre} \mid \tilde{s} \text{ is reachable from } \tilde{q}_0\}$.

Given a multi-agent system $M = (Q, Ag, \delta, q_0, \Pi, (\Pi_a)_{a \in Ag}, \pi)$, and an agent $a \in Ag$, we say that $M$ is $a$-**distinguished** if $\Gamma_a^M$ (relation defined on page 5) is a **congruence relation**, that is, an equivalence relation with the following property:

for any $q, r \in Q$, if $(q,r) \in \Gamma_a^M, (q,q') \in \delta, (r,r') \in \delta$ and
$$\pi_a(q') = \pi_a(r'), \text{ then } (q', r') \in \Gamma_a^M. \quad (5)$$

We utilize from now on the notation $\Gamma_a$ whenever the system $M$ is understood from the context.

PROPOSITION 8. *1. For any multi-agent system $M$, $\Delta_a(M)$ is an in-splitting of $M$. We denote this in-splitting as $\Delta_{a,M}^{-1} : \Delta_a(M) \to M$. Whenever the system $M$ is clear from the context, we use the notation $\Delta_a^{-1}$ instead of $\Delta_{a,M}^{-1}$.*

*2. For any agent $a \in Ag$ we have that $\Delta_a(M)$ is $a$-distinguished.*

PROPOSITION 9. *For any multi-agent system $M$ and two agents $a, b \in Ag$ with $\Pi_a \subseteq \Pi_b$, if $M$ is $b$-distinguished, then $\Delta_a(M)$ is $b$-distinguished too.*

PROPOSITION 10. *For any multi-agent system $M$, the following diagram commutes iff $M$ is $a$-distinguished:*

$$\begin{array}{ccc} 2^Q & \xrightarrow{K_a^f} & 2^Q \\ t_M^{-1} \downarrow & & \downarrow t_M^{-1} \\ 2^{\mathsf{supp}(t_M)} & \xrightarrow{K_a} & 2^{\mathsf{supp}(t_M)} \end{array} \quad (6)$$

*The same holds if the pair $K_a/K_a^f$ is replaced with $P_a/P_a^f$.*

*Definition 5.* We say that the pair of epistemic operators $K_a/K_a^f$, resp. $P_a/P_a^f$, **commutes for** $M$ if the diagram 6 is commutative for the respective pair.

Proposition 10 gives the first restricted form which ensures the commutativity of diagram 2 for formulas of the epistemic $\mu$-calculus. The second restricted form in which the pair $K_a/K_a^f$ (resp. $P_a/P_a^f$) commutes for a system is stated as point 2 in the next proposition:

PROPOSITION 11. *Consider two multi-agent systems $M_i = (Q_i, Ag, \delta_i, q_0^i, \Pi, (\Pi_a)_{a \in Ag}, \pi_i)$ with $Q_i = \{1,\ldots,n_i\}$, $(i = 1,2)$, related by an in-splitting $\chi : M_1 \xrightarrow{Ins} M_2$, and define the tree mapping $\hat{\chi} : \mathsf{supp}(t_{M_1}) \xrightarrow{Ins} \mathsf{supp}(t_{M_2})$, where $\hat{\chi}(\varepsilon) = \varepsilon$ and $\hat{\chi}(xi) = \hat{\chi}(x) \cdot \chi(i)$, for any $x \in \mathsf{supp}(t_{M_1})$ and $i \in Q_1$. Then the following properties hold:*

*1. $\hat{\chi}$ is a tree isomorphism between $t_{M_1}$ and $t_{M_2}$ and $t_{M_2} \circ \hat{\chi} = \chi \circ t_{M_1}$.*

*2. For any closed formula $\phi$ of the epistemic $\mu$-calculus for which the diagram 2 commutes in the system $M_2$, the following property holds:*
$$\|\phi\|_{M_1} = t_{M_1}^{-1}(\chi^{-1}(\lceil \phi \rceil_{M_2}))$$

REMARK 12. *The previous proposition tells us that, for closed formulas of the epistemic $\mu$-calculus for which diagram 2 commutes in $M_2$, in the eventuality that the system $M_2$ needs to be replaced with a "larger" system $M_1$ (for reasons related with the "subset construction" that ensures the first type of commutativity of $K_a/P_a$), the validity of $\phi$ on the tree $t_{M_1}$ can be recovered from the set of states $\chi^{-1}(\lceil \phi \rceil_{M_2})$, through the inverse tree mapping $t_{M_1}^{-1}$.*

We have now the essential ingredients that ensure the decidability of the model-checking problem for the $\mu$-calculus of nonmixing epistemic fixpoints. The algorithm runs as follows: we proceed by constructing the $Q$-operator interpretations of the subformulas of $\phi$ on the given system $M$, in a bottom-up traversal of the syntactic tree $T_\phi$. As long as we only treat subformulas not containing any epistemic operator, Theorem 3 ensures that these boolean operators are correct finitary abstractions of the tree semantics of our subformulas.

The first time we encounter in $T_\phi$ an epistemic operator, say, $K_a$, s.t. the subformula in the current node is $K_a \phi'$, we need to replace $M$ with its $a$-distinction, $\Delta_a(M)$, in order for the appropriate diagram to commute. This replacement is easier when $\phi'$ is a closed plain $\mu$-calculus formula. By combining Propositions 11 and 10, the tree semantics of the formula $K_a \phi'$ can be computed using the boolean operator $K_a^f(\Delta_a^{-1}(\lceil \phi' \rceil_M))$ in $\Delta_a(M)$, where $\Delta_a^{-1}(\lceil \phi' \rceil_M)$ represents the set of states in $\Delta_a(M)$ on which $\phi'$ holds.

The procedure is different when $\phi'$ is not closed. In this situation, we cannot determinize $M$, as observed in the remark 7. Therefore we need to descend along the syntactic tree to *all* the "nearest" nodes whose formulas are closed, and only there apply the $a$-distinction construction, as required by Proposition 11.

Suppose even further that $\phi'$ itself contains other knowledge operators, and some other knowledge operator $K_b$ is encountered during this descent. The "nonmixing" assumption on our formula implies that this other agent $b$ has compatible observability with our $a$ ($K_a$ and $K_b$ are not closed at the node associated with $K_a$). Therefore, the $a$-distinction of the models applied at lower levels commutes with $K_b$, fact which is ensured by Proposition 10 when the two agents have compatible observability.

This whole process ends when we arrive in the root of the syntactic tree, with an in-splitting $M'$ of the initial system $M$ and a (constant) boolean operator $\sigma$, which gives the finitary abstraction of the set of nodes of the tree $t_M$ where $\phi$ holds. The following paragraphs formalize this process.

PROOF OF THEOREM 4. Given a formula $\phi$ in the $\mu$-calculus of non-mixing epistemic fixpoints and a multi-agent system $M$, we associate with each node $x$ of $T_\phi$ an in-splitting mapping, denoted $T_\phi^{Ins}(x)$, such that the following properties hold:

1. For the root $\epsilon$ we have $T_\phi^{Ins}(\epsilon) = id_M$. Also for any non-closed node $x$ in $\mathsf{supp}(T_\phi)$, we have that $T_\phi^{Ins}(x) = id_{M'}$, where $M'$ is an in-splitting of $M$.

2. For any $x, xi \in \mathsf{supp}(T_\phi), i \in \{1,2\}$, $codom(T_\phi^{Ins}(x)) = dom(T_\phi^{Ins}(xi))$,

3. For any nodes $x_1, x_2 \in \mathsf{supp}(T_\phi)$ with $x_1 \leq x_2$, define first the *in-splitting mapping* between $x_1$ and $x_2$ as:
$$T_\phi^{Ins}(x_1...x_2) = T_\phi^{Ins}(x_1) \circ ... \circ T_\phi^{Ins}(x_2)$$
Then, for any leaves $x_1, x_2$ in $T_\phi$ we have that $T_\phi^{Ins}(\epsilon...x_1) = T_\phi^{Ins}(\epsilon...x_2)$, where $\epsilon$ is the root of $T_\phi$.

4. For any node $x_1$ which is a nearest closed successor of the root $\epsilon$, if $AgNCl_\phi(\epsilon) = \{a_1,...,a_k\}$ and $\Pi_{a_1} \subseteq ... \subseteq \Pi_{a_k}$, then $T_\phi^{Ins}(x_1)$ has the form:
$$T_\phi^{Ins}(x_1) = \Delta_{a_1}^{-1} \circ ... \circ \Delta_{a_k}^{-1} \circ \chi, \text{ for some } \chi,$$

Assuming that $T_\phi^{Ins}$ is constructed with all the properties above, we denote $InS(T_\phi^{Ins}) = T_\phi^{Ins}(\epsilon...x)$ where $x$ is any leaf in $T_\phi$. In the sequel, whenever we want to emphasize a property of the root of the syntactic tree $T_\phi$, we denote it $\epsilon^\phi$.

The construction of $T_\phi^{Ins}$ proceeds by structural induction on $\phi$. For the base case $\phi = p$ or $\phi = \neg p$, we put $T_p^{Ins}(\epsilon) = T_{\neg p}^{Ins}(\epsilon) = id_M$, for any $p \in \Pi$. Also for $\phi = Z, Z \in \mathcal{Z}$, note that, by construction, the root of $T_Z$ has a leaf successor which is the only child node. Then, $T_Z^{Ins}(\epsilon) = T_Z^{Ins}(1) = id_M$.

For the induction case, take a formula $\phi = Op.\phi'$ where $Op \in \{AX, EX, \mu Z, \nu Z\}$, and assume $T_{\phi'}^{Ins}(x)$ is defined. Then we put $T_\phi^{Ins}(1x) = T_{\phi'}^{Ins}(x)$ for any node $x$ of $\mathsf{supp}(T_{\phi'})$, and $T_\phi^{Ins}(\epsilon^\phi) = id_{M'}$, where $M' = dom(T_{\phi'}^{Ins}(\epsilon^{\phi'}))$.

Suppose $\phi = K_a \phi'$ or $\phi = P_a \phi'$. Note that for each node $1x$ which is not closed in $T_\phi$, the node $x$ is not closed in $T_{\phi'}$ either. Then we put $T_\phi^{Ins}(1x) = T_{\phi'}^{Ins}(x) = id_{M'}$, with $M'$ the appropriate multi-agent system. We also put $T_\phi^{Ins}(\epsilon^\phi) = id_{M_0}$ for the appropriate $M_0$. Furthermore, for each closed node $1x_1 \in \mathsf{supp}(T_\phi)$ which is *not* a nearest closed successor of $\epsilon^\phi$, we put $T_\phi^{Ins}(1x_1) = T_{\phi'}^{Ins}(x_1)$.

Take further a node $1x_1$ which is a nearest closed successor of the root $\epsilon^\phi$ and assume $AgNCL(\epsilon^\phi) = \{a_1,...,a_k\}$. By the above property 4 in the induction hypothesis, the in-splitting mapping in $x_1$ is $T_{\phi'}^{Ins}(x_1) = \Delta_{a_1}^{-1} \circ ... \circ \Delta_{a_k}^{-1} \circ \chi$ with $\Pi_{a_1} \subseteq ... \subseteq \Pi_{a_k}$. On the other hand, by the assumption that $\phi$ is a nonmixing formula, $a$ must have compatible observability with all the agents $a_1,...,a_k$. Therefore, there must exist some $i \leq k$ such that $\Pi_{a_1} \subseteq ... \subseteq \Pi_{a_i} \subseteq \Pi_a \subseteq \Pi_{a_{i+1}} \subseteq ... \subseteq \Pi_{a_k}$. We then define
$$T_\phi^{Ins}(1x_1) = \Delta_{a_1}^{-1} \circ ... \circ \Delta_{a_i}^{-1} \circ \Delta_a^{-1} \circ \Delta_{a_{i+1}}^{-1} \circ ... \circ \Delta_{a_k}^{-1} \circ \chi$$

Note that the domain and the codomain of each $\Delta_{a_j}^{-1}, (j \leq i)$ are different in $T_\phi^{Ins}$ from those in $T_{\phi'}^{Ins}$, due to the insertion of $\Delta_a^{-1}$.

According to the above constructions for $\phi = K_a\phi'$ of $\phi = P_a\phi'$, all the four properties are satisfied by $T_\phi^{Ins}$, the fourth one resulting from the construction of the in-splitting mapping for the nearest closed successors of the root.

Finally, take $\phi = \phi_1 Op \phi_2$ ($Op \in \{\wedge, \vee\}$). If $T_{\phi_1}^{Ins} = T_{\phi_2}^{Ins}$, put $T_\phi^{Ins}(1x) = T_{\phi_1}^{Ins}(x)$ for all nodes $x \in \mathsf{supp}(T_{\phi_1})$, $T_\phi^{Ins}(2x) = T_{\phi_2}^{Ins}(x)$ for all $x \in \mathsf{supp}(T_{\phi_2})$ and $T_\phi^{Ins}(\epsilon) = id_M$.

Suppose now $T_{\phi_1}^{Ins} \neq T_{\phi_2}^{Ins}$. Consider $AgNCl(1) = \{a_1,...,a_k\}$ and $AgNCl(2) = \{b_1,...,b_l\}$ with $\Pi_{a_1} \subseteq ... \subseteq \Pi_{a_k}$ and $\Pi_{b_1} \subseteq ... \subseteq \Pi_{b_l}$. Take then a node $x_1$ which is a nearest closed successor of the root of $T_{\phi_1}$, $\epsilon^{\phi_1}$, and a node $x_2$ which is a nearest closed successor of $\epsilon^{\phi_2}$. By the induction hypothesis we have:

$$T_{\phi_1}^{Ins}(x_1) = \Delta_{a_1}^{-1} \circ ... \circ \Delta_{a_k}^{-1} \circ \chi_1 \quad InS(T_{\phi_1}^{Ins}) = T_{\phi_1}^{Ins}(x_1) \circ \chi_1'$$
$$T_{\phi_2}^{Ins}(x_2) = \Delta_{b_1}^{-1} \circ ... \circ \Delta_{b_l}^{-1} \circ \chi_2 \quad InS(T_{\phi_2}^{Ins}) = T_{\phi_2}^{Ins}(x_2) \circ \chi_2'$$

with appropriate in-splittings $\chi_1, \chi_1', \chi_2, \chi_2'$.

On the other hand, by the assumption on $\phi$ being nonmixing, for any $i \leq k, j \leq l$, the two agents $a_i$ and $b_j$ must have compatible observability. It therefore follows that there exists a reordering of the union $\{a_1,...,a_k\} \cup \{b_1,...,b_l\}$ as $\{c_1,...,c_m\}$ such that $\Pi_{c_i} \subseteq \Pi_{c_{i+1}}$ for all $i \leq m-1$. Denote then:
$$\chi_0 = \Delta_{c_1}^{-1} \circ ... \circ \Delta_{c_m}^{-1}$$

By Proposition 9, $\chi_0$ is a $c$-distinction for any $c \in \{a_1,...,a_k\} \cup \{b_1,...,b_l\}$. Also, by property 2 of the induction hypothesis, $\chi_0$ is independent of the choice of the nodes $x_1, x_2$.

The same property from the induction hypothesis also ensures that, for any nearest closed successor $\overline{x}_2$ of $\varepsilon^{\phi_2}$, there exist in-splittings $\overline{\chi}_2^{\phi_2, \overline{x}_2}, \tilde{\chi}_2^{\phi_2, \overline{x}_2}$ such that:

$$T_{\phi_2}^{Ins}(\overline{x}_2) = \Delta_{b_1}^{-1} \circ ... \circ \Delta_{b_l}^{-1} \circ \overline{\chi}_2^{\phi_2, \overline{x}_2} \quad (7)$$
$$InS(T_{\phi_2}^{Ins}) = T_{\phi_2}^{Ins}(\overline{x}_2) \circ \tilde{\chi}_2^{\phi_2, \overline{x}_2} \quad (8)$$

We will then construct $T_\phi^{Ins}(\cdot)$ as follows:

1. For each closed node $x$ which is a leaf in $T_{\phi_1}$ but not a nearest closed successor of $\epsilon^{\phi_1}$, we put $T_\phi^{Ins}(1x) = T_{\phi_1}^{Ins}(x) \circ \chi_2 \circ \chi_2'$.

2. For each non-leaf, closed node $x$ in $T_{\phi_1}$ which is not a nearest closed successor of $\epsilon^{\phi_1}$ we copy $T_\phi^{Ins}(1x) = T_{\phi_1}^{Ins}(x)$.

3. For each nearest closed successor $x$ of $\epsilon^{\phi_1}$ which is not a leaf in $T_{\phi_1}$ we put $T_\phi^{Ins}(1x) = \chi_0 \circ \chi_1$.

4. For each closed node $x$ which is a leaf in $T_{\phi_1}$ and a nearest closed successor of $\epsilon^{\phi_1}$, we put $T_\phi^{Ins}(1x) = \chi_0 \circ \chi_1 \circ \chi_1' \circ \chi_2 \circ \chi_2'$.

5. For each closed node $x$ which is not a nearest closed successor of $\epsilon^{\phi_2}$ we copy $T_\phi^{Ins}(2x) = T_{\phi_2}^{Ins}(x)$.

6. For each closed node $x$ which is a nearest closed successor of $\epsilon^{\phi_2}$ we put $T_\phi^{Ins}(2x) = \chi_0 \circ \chi_1 \circ \chi_1' \circ \overline{\chi}_2^{\phi_2, x}$, where $\overline{\chi}_2^{\phi_2, x}$ is the in-splitting mapping associated with the node $x$ as in Identity 8 above.

7. For the root $\epsilon$ and the non-closed nodes $x$ of $T_\phi$, $T_\phi^{Ins}(\epsilon) = id_{M'}$ and $T_\phi^{Ins}(x) = id_{M''}$, with $M'$ and $M''$ appropriate multi-agent systems.

It's not difficult to see that the resulting mapping $T_{\phi_2}^{Ins}(\cdot)$ satisfies the five desired properties. More specifically, property 2 amounts to the following identity:

$$InS(T_\phi^{Ins}) = \chi_0 \circ \chi_1 \circ \chi_1' \circ \chi_2 \circ \chi_2'$$

Now we may show how $T_\phi^{Ins}$ can be used to build our algorithm. Let $M_x$ denote the multi-agent system which is the *domain* of the in-splitting $T_\phi^{Ins}(x)$, and denote $Q_x$ its state-space. Also, for convenience, we denote $\overline{M}_x$ the multi-agent system which represents the *codomain* of $T_\phi^{Ins}(x)$, and $\overline{Q}_x$ its state-space. Note that when $x, x1 \in \mathsf{supp}(T_\phi)$, $\overline{M}_x = M_{x1}$, and similarly $\overline{M}_x = M_{x2}$ when $x2 \in \mathsf{supp}(T_\phi)$.

Once we built the tree $T_\phi^{Ins}$, we associate with each node $x$ in $T_\phi$ a $\overline{Q}_x$-operator that will give all the information on the satisfiability of $form(x)$ in the given model. Formally, we build the tree $T_\phi^{str}$ whose domain is $\mathsf{supp}(T_\phi) \setminus \{x \mid T_\phi(x) = \top\}$ and which associates with each node $x$ a $\overline{Q}_x$-operator $T_\phi^{str}(x) : (2^{Q_x})^n \to 2^{Q_x}$. The construction will be achieved such that

$$\|form(x)\| \circ \left(t_{M_x}^{-1}\right)^n = t_{M_x}^{-1} \circ T_\phi^{str}(x) \qquad (9)$$

for each node $x$ with $form(x) \neq \top$.

The construction proceeds bottom-up on $\mathsf{supp}(T_\phi)$. We actually build *two* trees, $T_\phi^{str}$ and $\overline{T}_\phi^{str}$, such that $\overline{T}_\phi^{str}(x) : (2^{\overline{Q}_x})^n \to 2^{\overline{Q}_x}$ and $T_\phi^{str}(x) = \overline{T}_\phi^{str}(x) \circ \left[\left(T_\phi^{Ins}(x)\right)^{-1}\right]^n$, that is,

$$T_\phi^{str}(x)(S_1, \ldots, S_n) = \overline{T}_\phi^{str}(x)\left(\left(T_\phi^{Ins}(x)\right)^{-1}(S_1, \ldots, S_n)\right) \qquad (10)$$

Note that, once we build $\overline{T}_\phi^{str}(x)$ for a node $x$, $T_\phi^{str}(x)$ is defined by Identity 10, so we only explain the construction for $\overline{T}_\phi^{str}(x)$.

For nodes $x$ that are leaves in $T_\phi$ with $T_\phi(x) = p \in \Pi$, we put $\overline{T}_\phi^{str}(x) = \lceil p \rceil_M$, the constant $\overline{Q}_x$-operator. Recall that we do not define $T_\phi^{str}(x)$ for $\overline{T}_\phi(x) = \top$.

For $T_\phi(x) = Z_i \in \mathcal{Z}$ we put $\overline{T}_\phi^{str}(x)(S_1, \ldots, S_n) = S_i$, the $i$-th projection on $(2^{\overline{Q}_x})^n$.

For nodes $x$ with $T_\phi(x) = Op \in \{AX, EX, K_a, P_a \mid a \in Ag\}$ we put

$$\overline{T}_\phi^{str}(x)(S_1, \ldots, S_n) = Op^f\left(T_\phi^{str}(x1)(S_1, \ldots, S_n)\right)$$

For $T_\phi(x) = \wedge$ we put

$$\overline{T}_\phi^{str}(x)(S_1, \ldots, S_n) = \\ \left(T_\phi^{str}(x1)(S_1, \ldots, S_n)\right) \cap \left(T_\phi^{str}(x2)(S_1, \ldots, S_n)\right)$$

and similarly for $T_\phi(x) = \vee$, with $\cap$ replaced with $\cup$ in the above formula defining $\overline{T}_\phi^{str}(x)(S_1, \ldots, S_n)$.

For $T_\phi(x) = \mu Z_i$ with $1 \leq i \leq n$ we put

$$\overline{T}_\phi^{str}(x) = \mathsf{lfp}_{[T_\phi^{str}(x1)]}^i$$

and, similarly, for $T_\phi(x) = \nu Z_i$ we define

$$\overline{T}_\phi^{str}(x) = \mathsf{gfp}_{[T_\phi^{str}(x1)]}^i$$

The validity of Identity 9 follows then from Propositions 10 and 11.

The final step consists in checking whether $q_0^\varepsilon \in T_\phi^{str}(\varepsilon)$, where $q_0^\varepsilon$ is the initial state in the multi-agent system $M_\varepsilon$ associated with the root of $T_\phi$. The result of this check gives the answer to the problem whether $\varepsilon \models \phi$ in $t_M$.

□

The following result follows from a similar result for LTLK from [26]. A self-contained proof can be found in [3]:

THEOREM 13. *The model checking problem for the $\mu$-calculus of non-mixing epistemic fixpoints is hard for non-elementary time.*

## 5. CONCLUSIONS AND COMMENTS

We have presented a fragment of the epistemic $\mu$-calculus having a decidable model-checking problem. We argued in the introduction that the decidability result does not seem to be achievable using tree automata or multi-player games. Two-player games with one player having incomplete information and with non-observable winning conditions from [7] do not seem to be appropriate for the whole calculus as they are only equivalent with a restricted type of combinations of knowledge operators and fixpoints. We conjecture that the formula $\nu Z\big(p \vee AX.P_a Z\big)$ is not equivalent with any (tree automaton presentation of a) two-player game with path winning conditions. Translating this formula to a generalized tree automaton seems to require specifying a winning condition on concatenations of finite paths in the tree with "jumps" between two identically-observable positions in the tree. This conjecture extends the non-expressivity results from [6] relating $ATL$ and $\mu - ATL$.

The second reason for which the above-mentioned generalization would not work comes from results in [9] showing that the satisfiability problem for CTL or LTL is undecidable with the concrete observability relation presented here. It is then expectable that if a class of generalized tree automata is equivalent with the $\mu$-calculus of non-mixing epistemic fixpoints, then that class would have an undecidable emptiness problem and only its "testing problem" would be decidable. Therefore, the classical determinacy argument for two-player games would not be translatable to such a class of automata.

## Acknowledgments

We are grateful to D. Guelev for his careful reading of earlier versions of this paper.